# A Qualitative Analysis of Haptic Feedback in Music Focused Exercises

G. W. Young
Dept. Computer Science / Music
University College Cork
Cork, Ireland
g.young@cs.ucc.ie

D. Murphy
Dept. Computer Science
University College Cork
Cork, Ireland
d.murphy@cs.ucc.ie

J. Weeter
Dept. Music
University College Cork
Cork, Ireland
j.weeter@ucc.ie

## ABSTRACT
We present the findings of a pilot-study that analysed the role of haptic feedback in a musical context. To closely examine the role of haptics in Digital Musical Instrument (DMI) design an experiment was formulated to measure the users' perception of device usability across four separate feedback stages: fully haptic (force and tactile combined), constant force only, vibrotactile only, and no feedback. The study was piloted over extended periods with the intention of exploring the application and integration of DMIs in real-world musical contexts. Applying a music orientated analysis of this type enabled the investigative process to not only take place over a comprehensive period, but allowed for the exploration of DMI integration in everyday compositional and explorative practices. As with any investigation that involves creativity, it was important that the participants did not feel rushed or restricted. That is, they were given sufficient time to explore and assess the different feedback types without constraint. This provided an accurate and representational set of qualitative data for validating the participants' experience with the different feedback types they were presented with.

## ACM Classification
H.5.2 [Information Interfaces and Presentation] User Interfaces, Haptic I/O, Auditory (non-speech) feedback H.5.5 [Information Interfaces and Presentation] Sound and Music Computing

## 1. INTRODUCTION
Presented is an examination of device feedback executed in structured case-by-case studies. For analysis, four separate feedback types were explored in the performance of musical tasks and free-play. The effects of feedback were observed and recorded in note selection, melody following, and other explorative exercises. These tasks were selected to measure the perceived usability of the different feedback stages when presented in a musical performance context. By choosing this method of evaluation, it was possible to explore a qualitative approach in the evaluation of haptic DMIs applied in a creative venture; with focus remaining on the issues as examined in other studies of this type [1] [2] [3] [4].

Measuring a participant's experiences when playing music or working creatively with a DMI has been highlighted as being a highly complex operation that is often executed idiosyncratically [5] [6]. However, the formal evaluation of experience over time has been validated by studies in the field of Human-Computer Interaction (HCI) and Music [7] [8] [9]. Through the application of HCI evaluation tools in creative activities, the importance of *learnability* and *explorability* in structured evaluations has been identified as requiring extended periods of time to assess [10]. In addition, it is apparent that many DMI evaluations do not afford the participant adequate time to explore and evaluate these aspects of performance. Therefore, by incorporating the issues that longitudinal approaches are adept to exposing, an experiment was devised to investigate the experiences of musicians with device feedback applied in musical exercises.

## 2. BACKGROUND
Fundamentally, acoustic musical instruments convey performance information to musicians in the form of audio, visual, and haptic stimulation. In addition to this, the musician makes use of their own awareness of the relative position of the body and the strength of effort being employed in the interaction. The physical properties of sound generation in acoustic instruments causes the interface to return information in sympathy to the gestures applied to them. This information qualifies as information feedback, creating a tightknit relationship between the instrument and the performer (see figure 1). By combining both tactile with kinaesthetic stimulation, haptic feedback can be returned to the user, allowing for increased control in musical gesture articulation. Many new interfaces for musical expression require little or no direct contact with the gesture interface, returning no physical feedback to the user. Moreover, sound generation in current DMI designs is dealt with separately, divorcing musician from instrument, and failing to close the interaction feedback loop.

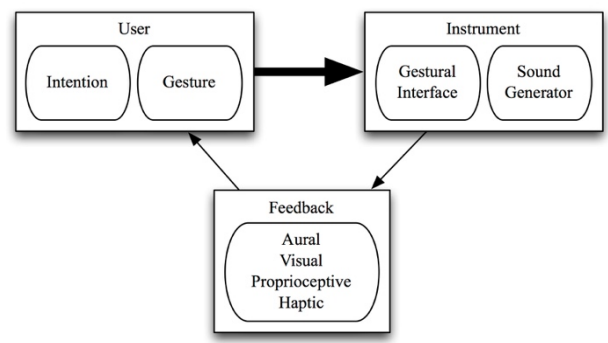

**Figure 1: Information feedback in Haptic Designs.**

DMIs that require no physical contact are often controlled via hand gestures, which are captured and relayed as data for the control of some synthesis parameters within an external audio synthesis engine. Bodiless and open-air instruments make use of video cameras and motion capturing (MOCAP) software to manipulate the synthesis parameters of the audio engine [11] [12]. Methods of noncontact gesture capture include ultrasonic or infrared sensors contained within a central transmitter [13] [14]. The capture of small, nuanced movements with no physical feedback present the NIME community with interesting performance and design challenges. The performer relies upon visual and proprioceptive feedback relating to their body position along with the audio response of the sound generator to their actions. This is adequate for most applications, but it has been observed that performers who have mastered their instrument also make use of haptic feedback cues in their performance [15]. Additionally, instruments that lack haptic feedback also present a *disconnect* between performer and device, creating a sense of loss in the sound produced and how it relates to movement [16].

Current analysis techniques have been successful in their appraisal of device feedback in task-based evaluations; however, they have arguably provided some inconclusive outcomes pertaining to the perception of usability relating to the specific types of feedback applied. This underpins the requirement for alternative methods of DMI evaluation applied in Computer Music and that understanding the user's experience of usability in a creative context is more complex than in traditional HCI evaluations. For example, the 'third paradigm' and the potential of haptics to introduce device embodiment [17]. While the implementation of quantitative feedback interaction analyses has largely been successful, the outputted data has arguably failed to account for the process of interaction or given any clear evidence of context-in-use affecting the participants' perception of usability. Therefore, contextual data should also be explored and holistic evaluations should also be required for the accurate evaluation of DMI feedback in creative musical applications.

## 3. ANALYSIS OF FEEDBACK

In the explorative study presented here, the users' experiences of device usability were collected to counteract the difficulties observed in evaluating feedback functionality in non-music contexts. Although the number of participants in many pilot studies would be considered too low for statistical analyses, a visual examination of the data can be used to highlight practical significances between the different feedback stages. Further to this, the data gathered can be later compared to both statistical and practical variations observed in previous studies. As an additional indication of these factors, experiments may also be designed to provide a structured and extensive period for the participants to adequately and accurately judge multiple factors.

The main goal of the presented pilot-study was to acquire information relating to the application of feedback in creative and explorative tasks by focusing on how the participants integrated the DMIs into a creative working process. As the appraisal of an individual's creativity and musicality is arguably subjective, the user's proficiency in composition and skill in the execution of musical tasks was not assessed; however, each participant self-evaluated their own performance with the device. Therefore, the participants were required to consider the strengths and weaknesses of the input metaphors at the different feedback stages and attempt to personalize their application and performance style to suit. This style of analysis was selected as a more qualitative analyses approach and address the requirements for a music-based analysis, as discussed in [2]. The perception of usability and the user's experiences when applying the different feedback types were recorded and then analysed via critical incident technique (CIT) analysis.

## 3.1 Device Descriptions

To analyse the role of haptic feedback in musical DMI interactions, two prototype devices were investigated. Each device was designed specifically to represent DMIs with a variety of feedback capabilities. These two devices also afford the user freedom of movement in a three-dimensional space around the device.

### 3.1.1 The Haptic Bowl

The Haptic Bowl is an isotonic, zero-order, alternative controller that was developed from a console game interface. The internal mechanisms of a Mad Catz "GameTrak" tethered spatial position controller were removed and relocated to a more robust and aesthetically pleasing shell. All original Human Interface Device (HID) circuitry was removed and replaced with an Arduino Uno smd edition. The HID upgrade reduced communication latencies and allowed for the expansion of device functionality through the addition of auxiliary buttons and switches. The controller has few movement restrictions as physical contact with the device is reduced to two tethers that connect the user via gloves. Control of the device requires the performer to visualize an area in three dimensions, with each hand tethered within this space.

### 3.1.2 The Non-Haptic Bowl

The Non-Haptic Bowl is also an isotonic, zero-order, alternative controller, based upon PING))) ultrasonic transducers and basic infrared (IR) motion capture (MOCAP) cameras. The components are arranged as digital inputs, via an Arduino Micro, and MOCAP cameras are attached via an integrated USB hub. The MOCAP system is crafted from modified Logitech C170 web cameras with internal IR filters removed and visual light filters covering the optical sensors. An IR light emitting diode (LED), embedded in a ring, is then used to provide a tracking source for the simple MOCAP system. The constituent components are contained within an aluminium shell, similar in size and shape as the Haptic Bowl. The use of these sensors best matched the input capabilities of the Haptic Bowl, ensuring a comparable interaction. However, the device has fewer movement restrictions than the Haptic Bowl, as no physical contact is required.

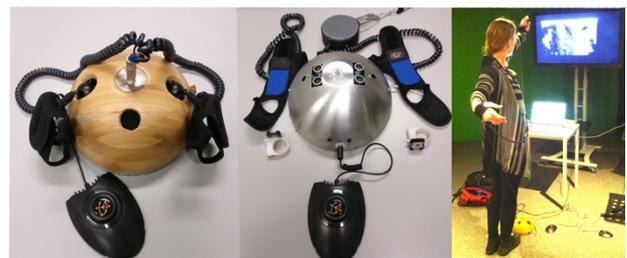

**Figure 2: The Haptic Bowl (left), Non-haptic Bowl (center), and User for Scale (left).**

### 3.1.3 Feedback Methodologies

In addition to the user's aural, visual, and proprioceptive awareness, haptic feedback components were incorporated into the DMIs to communicate performance data back to the user. In the Haptic Bowl, additional feedback was included in the form of strengthened constant-force return mechanisms for both tethers and audio frequency vibrotactile feedback delivered via actuators embedded in gloves. The audio-related vibrotactile feedback was supplied via a Bluetooth speaker embedded within the Haptic Bowl (a modified Logitech X100 Mobile Wireless Speaker) and connected via an audio connection on the top of the device to vibrotactile actuators contained within the Audio-Tactile Glove [18]. It was possible to apply audio frequency vibrotactile feedback to the Non-Haptic bowl via the same gloved actuators. For the Non-Haptic Bowl the audio output from the sound generator was routed to the same type of Bluetooth speaker, but it was kept external from the main device to demonstrate the disconnect of these feedback sources in current DMI designs. From combinations formulated around these feedback techniques, it was possible to create four feedback profiles:

1. Haptic feedback (both force and vibrotactile feedback)
2. Force feedback (force feedback only)
3. Tactile feedback (vibrotactile feedback only)
4. No feedback (no physical feedback)

All the above combinations operated within the predefined sensory requirements for haptic feedback, outlined by Berdea and Coiffet [19].

## 4. EXPERIMENT

Case studies took place individually in a sound proofed recording studio and lasted on average 4 to 5 hours in total. The USB output from each device was connected to a 2012 MacBook Pro Retina. The input data from the devices were converted in Processing into OSC messages and outputted as UDP information over port 12001. Pure Data (PD), an open source visual programming language, was used to receive serial data. Within PD, a polyphonic sound generator was programmed that incorporated variable pitch, amplitude, and an attack, decay, sustain, release envelope generator. The inputted gesture data was used to control each element of the sound generator as such: the right hand X/Y/Z input stage of the device controlled the parameters of attack, sustain, and pitch respectively; the left hand X/Y/Z input was used to control decay, release, and total volume.

### 4.1 Participants

Six musicians participated in the study. All participants were recruited from previous experiments and were therefore familiar with the input devices and their operation. The participants were aged 22 to 29 ($M = 24.5$, $SD = 2.69$) and consisted of 5 males and 1 female. All participants self-identified as being musicians, having been formally trained or regularly composing and performing Computer Music in the past five years.

### 4.2 Methodology

Participants were presented with each feedback type in counterbalanced order. Participants were then asked to perform each task in random order. Following this, each participant was interviewed to evaluate the performance of the feedback stage. The short experts of music and simple exercises were validated in the works of Orio and Wanderley [20], and O'Modhrain [21]. The sheet music was provided in advance of the session to give participants some familiarity, but exploration and free play was encouraged to investigate the potential application of feedback in a variety of different performance contexts. Each feedback type was evaluated independently after an appropriate period had passed, judged solely by the participant. This time varied, but was on average 60 to 90 minutes per feedback type. A post-task interview was completed for each feedback stage to expand upon the opinions expressed during the study. Participants were asked to perform the following tasks:

1. Generate isolated tones, from simple triggering to varying characteristics of pitch, loudness, and timbre.
2. Perform musical gestures specific to the device, such as glissandi, trills, grace notes, and so on.
3. Play simple scales and arpeggios altering speed, range, and articulation.
4. Repeat phrases with different contours and variations of structure.
5. Play continuous feature modulation (e.g. timbre, amplitude or pitch) both for a given note and inside a phrase.
6. Play simple rhythms at different speeds combining tones or pre-recorded material.

All interviews followed the same guiding question: *What were the central elements of device feedback that resulted in task success or failure?* This question was then operationalized by the following procedure:

- What positive attributes did the feedback display?
- What negative attributes did the feedback display?
- What features made the task a success or failure?
- Describe this success or failure in a musical context?

Interview-laddering was applied throughout to explore the subconscious motives that lead to a specific criterion being raised. A Content Analysis was then applied to extrapolate upon the interview data collected. This set of procedures was used to systematically identify any behaviours that contributed to the success (positive) or failure (negative) in the specific context.

### 5. RESULTS

From the interview-transcripts, coherent thoughts and single statements were identified and extracted. After redundancy checking, a final total of 93 comments were counted ($M = 23$, $SD = 1.9$; per feedback stage). Following this, three researchers were independently employed to iteratively classify this pool of statements as either 'positive' or 'negative' performance evaluations, as can be seen in Table 1. Although this process was reductive, further analysis of this data is expected to develop a bottom-up categorical system of classifications. Known areas of concern in musical interactions include *Learnability*, *Explorability*, *Feature Controllability*, and *Timing Controllability* [10].

In the application of longitudinal-style studies, all of the participants expressed an appreciation of post-execution, free-exploration of musical tasks. This was deemed important as all participants had expressed a wish to explore the compositional and improvisational capabilities of the DMI.

### 5.1 Haptic Feedback

For the first exercise, the participants expressed mixed opinions about the performance of the haptic feedback device. Both positive and negative attributes were identified as affecting performance, but overall, there were more negative attributes identified than positive. Basic musical gestures were thought as being easier at slower tempos and more difficult at faster measures. However, participants highlighted that with more practice they might be able to achieve more success and therefore the task would become easier. Simple scales and arpeggios were considered easy to perform, but only within the limitations of the sound generator. However, the participants were able to adapt the short musical excerpts to fit within the constraints of the DMI. In the manipulation of phrases, the participants were comfortable playing the music presented to them and easily modified it to fit their own performance styles. Initially, participants required both the sheet music and short audio clips to assist in performing the different phrases. However, once familiar, they could all confidently play without assistance. To evaluate the different features afforded, participants performed a variety of different styles of music. Furthermore, the participants introduced variations in the presented music to explore modulations in the performance of the scores. In the final category of musical exercises, the participants were keen to comment upon the possibility of varying the tempos of the presented scores and discuss the challenges presented when playing along with other sources of music. These comments were mainly positive, indicating that the feedback was influential in applications that required consideration of tempo.

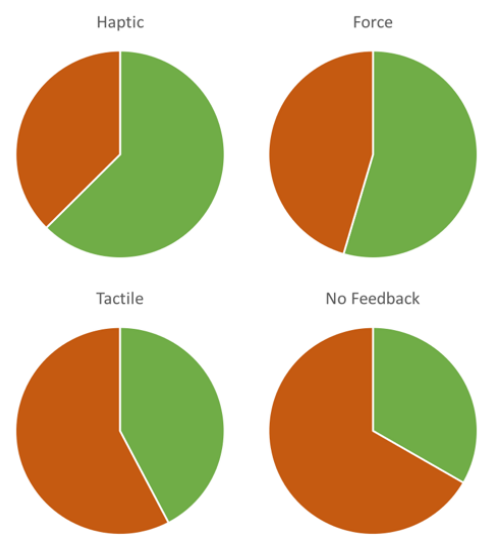

**Figure 3: Total success (green) and failure (red) comments.**

Table 1: The number of Positive (√) and Negative (X) Comments made for Tasks 1 to 6.

| Task<br>Feedback Type | 1 √ | 1 X | 2 √ | 2 X | 3 √ | 3 X | 4 √ | 4 X | 5 √ | 5 X | 6 √ | 6 X | Total |
|---|---|---|---|---|---|---|---|---|---|---|---|---|---|
| Haptic | 2 | 3 | 3 | 0 | 2 | 2 | 3 | 1 | 2 | 2 | 3 | 1 | 24 |
| Force | 2 | 2 | 3 | 0 | 1 | 1 | 2 | 1 | 3 | 3 | 1 | 3 | 22 |
| Tactile | 3 | 2 | 1 | 3 | 0 | 4 | 3 | 3 | 3 | 1 | 1 | 2 | 26 |
| No Feedback | 1 | 3 | 0 | 3 | 0 | 2 | 1 | 0 | 4 | 3 | 1 | 3 | 21 |
| Total Comments | 8 | 10 | 7 | 6 | 3 | 9 | 9 | 5 | 12 | 9 | 6 | 9 | 93 |

## 5.2 Force Feedback

For the force feedback stage, the control of single tones was evaluated both positively and negatively. The main issues with this feedback stage related to the participants' self evaluation of accuracy performance in pitch selection. However, all participants thought force feedback to be easier to use than the tactile and no feedback stages. There were very few definitive comments made about performing scales and arpeggios. This was initially attributed to the ease in which the participants completed the tasks. However, it was also observed that participants felt undecided about their own performance rather than that of the feedback in these actions. Phrases were generally seen to be easy to perform. However, the perception of comfort in task was dependent upon tempo. The continuous control of features was considered to be more difficult than the haptic feedback stage, with more time required in achieving an acceptable performance. There were mixed thoughts on the application of force feedback in exercises that required rhythms performed in combination with other materials.

## 5.3 Tactile Feedback

Controlling the different functions of the sound generator was done so enthusiastically; however, this behavior varied in form between participants. The control of the sound generator was perceived as being difficult. However, all participants preferred tactile feedback over no feedback. In the performance of scales and arpeggios, participants felt that accuracy was severely lacking. In general, all participants felt that movements requiring precision were difficult to achieve. Most participants considered the creation of phrases impartially for this feedback type. There were difficulties, but it was felt that they could be overcome with some training and practice. The continuous audio-related feedback was thought to be a positive feature for most tasks. The feedback was considered to assist in pitch and intensity precision; however, some of the other parameters were much harder to control. The application of the tactile feedback stage in the performance of rhythms was generally thought to be very difficult. However, in these particular tasks tactile feedback was also thought to be more advantageous than no feedback. The transparency of movement and perception of latency were both thought to be lacking for the control of other materials.

## 5.4 No Feedback

Participants considered controlling the different characteristics of the sound generator as being difficult with no feedback. However, further control might be achievable given more time with the device. A perceived reduction in the responsiveness of the device was expressed by all participants. Therefore, basic musical gestures were thought to be difficult to perform with any accuracy. The level of precision afforded to the users was considered much poorer with no feedback. Thus, scales and arpeggios were very difficult to complete. Furthermore, as accurate control perception was not achieved, some of the participants could not make coherent or congruent statements about performing phrase contours. However, one participant thought that they were somewhat achievable. In the control of continuous features of the sound generator, the participants expressed a requirement for more time and practice. It was found that although control was achievable, it took a longer time in comparison to the other feedback stages. When performing rhythms with no feedback, the participants preferred slower tempos over faster. In addition, when controlling different parameters, the participants struggled to perceive scenarios where the precise control of external parameters could be achieved.

## 6. DISCUSSION

Within the field of Computer Music, audio-visual interface devices dominate commercial markets and haptic feedback is neglected or presented as a novel feature in a device's interaction methodology. Examples of this can be seen in USB piano keyboards, basic slider and button controllers, and many of the digital renditions of interactive instruments and sequencers that are available as downloadable applications on touch-screen mobile devices. The results of the analyses presented in this study have suggested that there is a potential to improve upon a user's experience and increase the capacity of information that can be physiologically communicated in interactions that include haptic feedback. In addition to this, the results of the experiments also suggest that neglecting feedback in a DMI's design has a negative effect upon aspects of a user's perception of device accuracy.

Comparisons between functionality testing and the explorative case studies also highlight important factors of consideration in evaluating the successful completion of a musical task versus a less constrained creative endeavour, as a DMI cannot simply be determined as usable without context. Instead, it was observed how a DMI applied as a tool and the experiences of creativity can be used as the composite of several qualities that are heuristically discovered and determined by the artist. Therefore, it can be concluded that the usability of a DMI should not be analysed alone or outside of the context of a specific application. Usability can instead be better understood as a factor of user experience that emphasises the importance of the context of an evaluation, whereas the overall user experience should serve to quantify the factors that influence a user's application of specific technologies. Particularly to the findings presented within this study, regard to the senses involved in a musical interaction can be considered as a highly influential factor on user experience. However, it should also be acknowledged that touch is understandably reduced in importance below that of aural in the musical interactions witnessed.

In HCI, a usability analysis seeks to quantify an interaction between a user and a device in a specific way to ascertain if the device is proficient in undertaking the tasks it was designed for. In music, musicians perform a similar evaluation when appraising the potential of an instrument before composing for or performing. However, as there is currently no specific questionnaire designed for musical task analysis [1], and in the case presented, an analysis of discourse was used as a precursor to the design and creation of one. Although there were some variations in the number of comments made, there was also clear preference for haptic feedback over the other feedback stages in the exercises performed and analysed. Further to this, force feedback was rated more favourably over tactile. However, other feedback design configurations may have been explored here to reveal more specific data relating to the kinds of feedback applied and their effects upon the user, as was found in [22]. Furthermore, the observations presented in our experiment are based on a small number of participants and therefore care must be taken not to elaborate beyond the success/failure data presented. Also, although the data yielded several significant practical effects of feedback, there were no comparisons made to any quantifiable parameters as can be found in functionality focused experiments.

As mentioned, consideration is required of the number of participants who presented for this analysis, as it may raise questions of significance. In response, it is argued that the participants were afforded a much longer period of exploration and freedom of application than has previously been seen in function focused experiments presented at NIME. Additionally, the analysis concentrated on the performance of the feedback type and not the performance of the individual, compensating for subjectivity in device analysis. Furthermore, the significance of any differences in participant responses to feedback was concluded via visual observations in the data gathered. This analysis criterion was applied as statistical significance is recognised as difficult to determine for single-subject experimental procedures. Single-subject conditioning studies are common in physiological and psychological measures; however, it can be argued that this method can be subjective and researcher biased results could potentially be observed. To counteract this effect, interpretation of the data gathered was done so comparatively based upon the same questions that have been observed in analyses of this type presented at NIME. Although previous device analyses in the evaluation of DMIs in musical contexts have been described as idiosyncratic, it is important to acknowledge that levels of statistical significance and practical (clinical) meaningfulness can also present quite differently in studies of this type. Therefore, by following a structured analysis methodology it was possible to reduce researcher interpretations of data, as care was taken to elaborate only upon significant differences with the same clarity of deduction and extraction of meaning and interpretation as has been reported in the supporting literature of the field.

## 7. CONCLUSIONS

In the study presented, the application of different feedback types in musical tasks presented with an observable advantage over no feedback. The analysis of participant responses revealed that there was a perceivable qualitative difference between feedback in the successful execution of musical exercises. The results revealed that the participants preferred device feedback in the order of haptic, force, and tactile over no feedback. This study will hopefully also provide a foundation of familiar language between quantitative and qualitative data, and examples of the practical application of design testing methodologies. It is therefore suggested that the inclusion of multiple performance factors should be a fundamental aspect of any rigorous device analysis.

## REFERENCES

[1] G. M. Schmid, "Measuring Musician's Playing Experience: Development of a questionnaire for the evaluation of musical interaction," in *New Interfaces for Musical Expression*, London, 2014.

[2] G. W. Young and D. Murphy, "HCI Paradigms for Digital Musical Instruments: Methodologies for Rigorous Testing of Digital Musical Instruments," in *Computer Music Multidisciplinary Research*, Plymouth, 2015.

[3] D. Stowell, A. Robertson, N. Bryan-Kinns and M. D. Plumbley, "Evaluation of Live Human–Computer Music-Making: Quantitative and qualitative approaches," *Human-Computer Studies,* vol. 67, no. 11, pp. 960-975, 2009.

[4] J. Barbosa, J. Malloch, M. M. Wanderley and S. Huot, "What does" Evaluation" mean for the NIME community?," in *New Interfaces for Musical Expression*, Baton Rouge, 2015.

[5] S. Gelineck and S. Serafin, "Longitudinal Evaluation of the Integration of Digital Musical Instruments into Existing Compositional Work Processes," *Journal of New Music Research,* vol. 41, no. 3, pp. 259-276, 2012.

[6] G. W. Young and D. Murphy, "Digital Musical Instrument Analysis: The Haptic Bowl," in *Computer Music Multidisciplinary Research*, Plymouth, 2015.

[7] G. M. Schmid, A. N. Tuch, S. Papetti and K. Opwis, "Three Facets for the Evaluation of Musical Instruments from the Perspective of the Musician.," in *ACM CHI*, San Jose, 2016.

[8] J. Barbosa, F. Calegario, V. Teichrieb, G. Ramalho and P. McGlynn, "Considering Audience's View Towards an Evaluation Methodology for Digital Musical Instruments," in *New Interfaces for Musical Expression*, Ann Arbor, 2012.

[9] J. J. Kaye, "Evaluating Experience-Focused HCI," in *CHI'07 extended abstracts on Human factors in computing systems*, New York, 2007.

[10] R. Poppe, R. Rienks and B. Van Dijk, "Evaluating the Future of HCI: Challenges for the evaluation of emerging applications," in *Artificial Intelligence for Human Computing*, Berlin, Heidelberg: Springer, 2007, pp. 234-250.

[11] N. Orio, N. Schnell and M. M. Wanderley, "Input Devices for Musical Expression: Borrowing tools from HCI," in *New Interfaces for Musical Expression*, Seattle, 2001.

[12] A. Hornof and L. Sato, "Eyemusic: Making music with the eyes," in *New Interfaces for Musical Expression*, Hamamatsu, 2004.

[13] K. Mase and T. Yonezawa, "Body, Clothes, Water and Toys: Media Towards Natural Music Expressions with Digital Sounds," in *New Interfaces for Musical Expression*, Seattle, 2001.

[14] D. Livingstone and E. R. Miranda, "Orb3: Adaptive interface design for real time sound synthesis & diffusion within socially mediated spaces," in *New Interfaces for Musical Expression*, Vancouver, 2005.

[15] R. Rich, "Buchla Lightning MIDI Controller: A powerful new midi controller is nothing to shake a stick at," *Electronic Musician,* vol. 7, no. 10, pp. 102-108, 1991.

[16] P. R. Cook, "Remutualizing the Musical Instrument: Co-design of synthesis algorithms and controllers," *New Music Research,* vol. 33, no. 3, pp. 315 -320, 2004.

[17] J. Rovan and V. Hayward, "Typology of Tactile Sounds and their Synthesis in Gesture-Driven Computer Music

Performance," in *Trends in Gestural Control of Music*, M. Wanderly and M. Battier, Eds., Editions IRCAM, 2000.

[18] G. Young, D. Murphy and J. Weeter, "Audio-Tactile Glove," in *Digital Audio Effects*, Maynooth, 2013.

[19] G. Berdea and G. Coiffet, Virtual Reality Technology, New York, NY: J. Wiley & Sons Inc., 1994, p. 4.

[20] N. Orio and M. M. Wanderley, "Evaluation of Input Devices for Musical Expression: Borrowing tools from HCI," *Computer Music Journal,* vol. 26, no. 3, pp. 62-76, 2002.

[21] M. S. O'Modhrain, "Playing by Feel: Incorporating haptic feedback into computer-based musical instruments," Stanford University, Stanford, 2001.

[22] E. Berdahl, J. O. Smith, S. Weinzierl and G. Niemeyer, "Force-sensitive detents improve user performance for linear selection tasks," *Transactions on Haptics,* vol. 6, no. 2, pp. 206-216, 2013.

[23] C. Kiefer, N. Collins and G. Fitzpatrick, "HCI Methodology for Evaluating Musical Controllers: A case study," in *New Interfaces for Musical Expression*, Genova, 2008.

[24] G. M. Schmid, "Evaluation of musical instruments from the musician's perspective: A questionnaire for assessing the musician's perception of the experiential qualities of musical instruments.," University of Basel, Basel, 2015.